\documentclass[a4paper,11pt]{article}

\usepackage{geometry}
\usepackage{fancyhdr}
\usepackage{amsmath, amssymb}
\usepackage{graphicx}
\usepackage{color}
\definecolor{bg}{rgb}{0.89, 0.95, 0.71} 
\definecolor{ac}{rgb}{0.1, 0.1, 0.9} 
\definecolor{rc}{rgb}{0.77, 0.57, 0.71} 
\usepackage{listings}
\lstdefinestyle{latex}{language=TeX,
                       backgroundcolor=\color{bg},
                       basicstyle=\small\ttfamily,
                       frame=leftline,
                       xleftmargin=1.4em,
                       framexleftmargin=.8em}
\lstdefinestyle{cmdline}{
                         }
\usepackage{url}
\usepackage[pdftex,
            bookmarks,
            colorlinks=false,
            linkcolor=black,
            plainpages = false,
            pdfpagemode = UseNone,
            pdfstartview = FitH,
            citecolor = green, urlcolor = cyan, filecolor = magenta]{hyperref}

\usepackage[para]{footmisc}
\makeatletter
\long\def\@makefntext#1{\leavevmode
\@makefnmark\nobreak
\hskip.05em\relax#1%
}
\makeatother

\newcommand{\bibliolink}[2]{\href{#1}{\nolinkurl{#1}}.}

\makeatletter 

\def\subsection{\@startsection{subsection}{2}{\z@}{1.5ex plus .3ex minus
   .1ex}{.2ex plus .1ex}{\hspace*{-3em}\bf\large\color{ac}}}
\makeatother
\newlength{\rulelength}
\makeatletter
\renewcommand{\@makecaption}[2]{%
		    \sbox{\@tempboxa}{{\small #1: #2}}
		    \ifdim \wd\@tempboxa > \hsize
		    {\small #1: #2}\par
		    \else
		    \global\@minipagefalse
		    \hbox to \hsize {\hfil {\small #1: #2}\hfil}%
		    \fi
		   \vspace{\belowcaptionskip}
		     }
\makeatother

\newcommand\mysecv[1]{{\bigskip\bf\large #1}}
\newcommand{\arten}[5]{\textit{#1}, #2 {\bf{#3}}, #4 (#5).}
\newcommand{\artenarx}[6]{\textit{#1}, #2 {\bf{#3}}, #4 (#5) #6.}

\begin{document}

\begin{center}
\thispagestyle{empty}

{\bf\large Possibility of Using a Satellite-Based Detector for Recording Cherenkov Light\\
from Ultrahigh-Energy Extensive Air Showers Penetrating into the Ocean Water}\\[4mm]
Olga P. Shustova\footnote{\nolinkurl{olga.shustova@eas.sinp.msu.ru}}\\[1mm]
{\it\small Faculty of Physics, Lomonosov Moscow State University,\\Leninskie gory 1/2, Moscow 119991, Russia}\\[4mm]
Nikolai N. Kalmykov\footnote{\nolinkurl{kalm@eas.sinp.msu.ru}}, Boris A. Khrenov\\[1mm]
{\it\small Skobeltsyn Institute of Nuclear Physics, Lomonosov Moscow State University,\\Leninskie gory 1/2, Moscow 119991, Russia}
\end{center}

\begin{abstract}
We have estimated the reflected component of Cherenkov radiation, which arises in developing of an~extensive air shower with primary energy of~$10^{20}$\,eV over the ocean surface. It has been shown that, under conditions of the TUS experiment, a f{}lash of the ref{}lected Cherenkov photons at the end of the f{}luorescence track can be identified in showers with zenith angles up to $\approx\!20^\circ$.
\end{abstract}

\mysecv{Introduction}

At present new space experiments, TUS\,\cite{tus} and JEM-EUSO\,\cite{jem_euso}, are being prepared to study ultrahigh-energy cosmic rays (UHECRs). Such an energetic particle generates in the atmosphere an extensive air shower (EAS) accompanied by huge f{}luxes of f{}luorescence and Cherenkov photons.

The f{}luorescence radiation is isotropic, which makes it possible to detect the tracks of the showers by detectors with a wide field of view at large distances. The first similar devices were the ground-based array F{}ly's Eye\,\cite{flys_eye} and its improved successor HiRes\,\cite{hires}. An idea of recording f{}luorescence tracks of EASs by a satellite in the Earth orbit was suggested in 1981 by R.\,Benson and J.\,Linsley\,\cite{Linsley}. The important advantage of space observations as compared to ground ones is the possibility of searching for UHECRs with one detector over the whole celestial sphere.

The Cherenkov radiation, on the contrary, is mainly concentrated in the direction of motion of the EAS electrons disk. Some part of the Cherenkov photons, however, may move upward due to atmospheric scattering and ref{}lecting from a high-albedo surface. For example, the balloon experiments\,\cite{balloon} have been proposed and now in active preparation for detecting the Cherenkov photons ref{}lected from the snow surface.

With a satellite-based detector spending two thirds of its operation time over the ocean, the possibility of recording the Cherenkov light from a shower penetrating into the ocean water is of particular interest. In this case several components of the Cherenkov radiation, moving into the upper hemisphere, can be distinguished. These are 1) atmospheric photons ref{}lected from the water surface, 2) photons created by the tail of the shower in the water and emerging to the surface due to scattering, 3) atmospheric photons coming into the water and emerging again to the surface. Additionally, a fraction of the atmospheric Cherenkov radiation moves upward due to scattering in air. Registration of the photons ref{}lected from the ocean surface at the end of the f{}luorescence track of the shower would enable to determine its trajectory more precisely.

In this work we perform calculations for an EAS with primary energy $E_0\!=\!10^{20}$\,eV and impose constraints on the shower zenith angles at which flashes of the reflected component can be detected by the TUS detector.

\mysecv{1. Calculation technique}

In this section we sketch out the calculation technique. Detailed description can be found in\,\cite{Shustova}.

We apply the Monte--Carlo method for generation of Cherenkov photons by electrons of a shower and their further propagation in a medium. During simulations the fate of a package of photons (hereafter, just a photon) is predicted. Each photon is characterized by a certain value of weight, i.e. the number of ``surviving" photons at a given moment with respect to the maximum possible number of photons at the point of their generation. As the laboratory frame we consider a system in which the origin $O$ is placed at the intersection point of the shower axis with the f{}lat ocean surface (the plane $Oxy$) and the $Oz$ axis is directed to the atmosphere boundary.

First we determine the position, direction and weight of a photon generated by a parent electron, which further are used as input parameters in the procedure of the photon propagation. Coordinates of the photon are defined by depth of the electron, simulated from the Gaisser--Hillas formula with parameters $N_\mathrm{max}\!=\!7\!\cdot\!10^{10}$, $X_\mathrm{max}\!=\!800$\,g/cm$^2$, $\lambda\!=\!80$\,g/cm$^2$ for the mean cascade curve of a shower with primary energy $E_0\!=\!10^{20}$\,eV. To specify directional cosines of the photon in the laboratory system we need to determine energy of the electron and its angles relative to the shower axis. Their values are simulated from corresponding distributions proposed in\,\cite{Lafebre}.

Then we follow the procedure of the photon propagation in air and ocean water. The length of travel, direction and weight of the photon are determined at every scattering event. In the case of air we take into account Rayleigh scattering, with the atmosphere assumed to be isothermal. Data on mean free paths of scattering and absorption in water are taken from\,\cite{Mobley}. The parameter value of the Henyey--Creenstein distribution used for simulating the scattering angle of the photon is also given there. For ref{}lection and refraction of the photon at the interface between two media are applied formulae of geometric optics.

Directional cosines and weight of the photon ref{}lected are determined with regard for a wavy surface of the ocean. We use simple model of a fixed wave, with the form described by parametric equations of a truncated cycloid. The wave front is parallel to the $Oy$ axis. The parameters, wavelength $\lambda_{w}\!=\!40$\,m and amplitude of the wave $h_{w}\!=\!0.7$\,m, are chosen on the basis of the satellite data\,\cite{wind_speed} and the wave distributions over amplitudes and half-wavelengths\,\cite{wave_parameters} and are in compliance with weather conditions favorable for recording EASs by a space-based detector.

As a result of these simulations we obtain an estimate for all components of the Cherenkov light, which occur in propagating of a shower over the water surface. As we have expected, the numbers of ref{}lected and scattered atmospheric photons in relation to the total number of photons produced in the air are the largest in magnitude. For instance, they amount to $\approx\!1.5$\,\% for showers with zenith angles up to $\approx\!30^\circ$ and for photons with wavelengths within the range of $300-400$\,nm.

\mysecv{2. Signal estimation under the TUS detector conditions}

Here we estimate the magnitude of the Cherenkov signal that can be recorded by the TUS detector with the following characteristics. The mean height of the satellite orbit is assumed to be 400\,km. The field of view and the mirror area of the detector are $\approx\!9.5^\circ$ and 1\,m$^2$, respectively. The photosensor is a matrix of $16\!\times\!16$ cells of a photoelectronic multiplier (PEM), with the field of view of 0.005 rad ($\approx\!29^\circ$) for a single cell.

\begin{figure}[h!]
\begin{center}
\includegraphics[height=15pc]{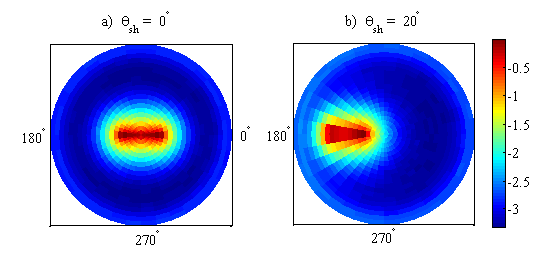}
\end{center}
\vspace{-1.5pc}
\caption{Angular distribution of ref{}lected Cherenkov photons near the ocean surface for showers with angles $\varphi_\mathrm{sh}\!=\!0^\circ$ and $\theta_\mathrm{sh}\!=\!0^\circ$\,(a), $20^\circ$\,(b). The surface density of the photons is projected onto a circle with unit radius. The circle is divided into cells with area $\Delta r\Delta\varphi\sin\theta$, where $\Delta r\!=\!0.05$ and $\Delta\varphi\!=\!\pi/18$. The direction of a photon is characterized by angles $\theta$ and $\varphi$. A value of $r$ along the radius of the circle determines $\sin\theta$. Numerals denote values of $\varphi$. The scale corresponds to the decimal logarithm of the relative density of photons.}\label{pic1}
\end{figure}

Let us consider the case of a vertical shower with primary energy $E_0\!=\!10^{20}$\,eV. Fig.\,\ref{pic1}(a) illustrates the angular distribution of the ref{}lected Cherenkov component near the ocean surface. As can be seen, considering waves on the ocean surface extends the range of the photon zenith angles $\theta$ across the wave front. The TUS detector will be able to record the photons hitting the mirror, on condition that $\theta\!\leqslant\!4.75^\circ$. In Fig.\,\ref{pic1} these photons are bounded by the central circle with radius $r\!\approx\!0.083$.

\begin{figure}[h!]
\begin{center}
\includegraphics[height=25pc]{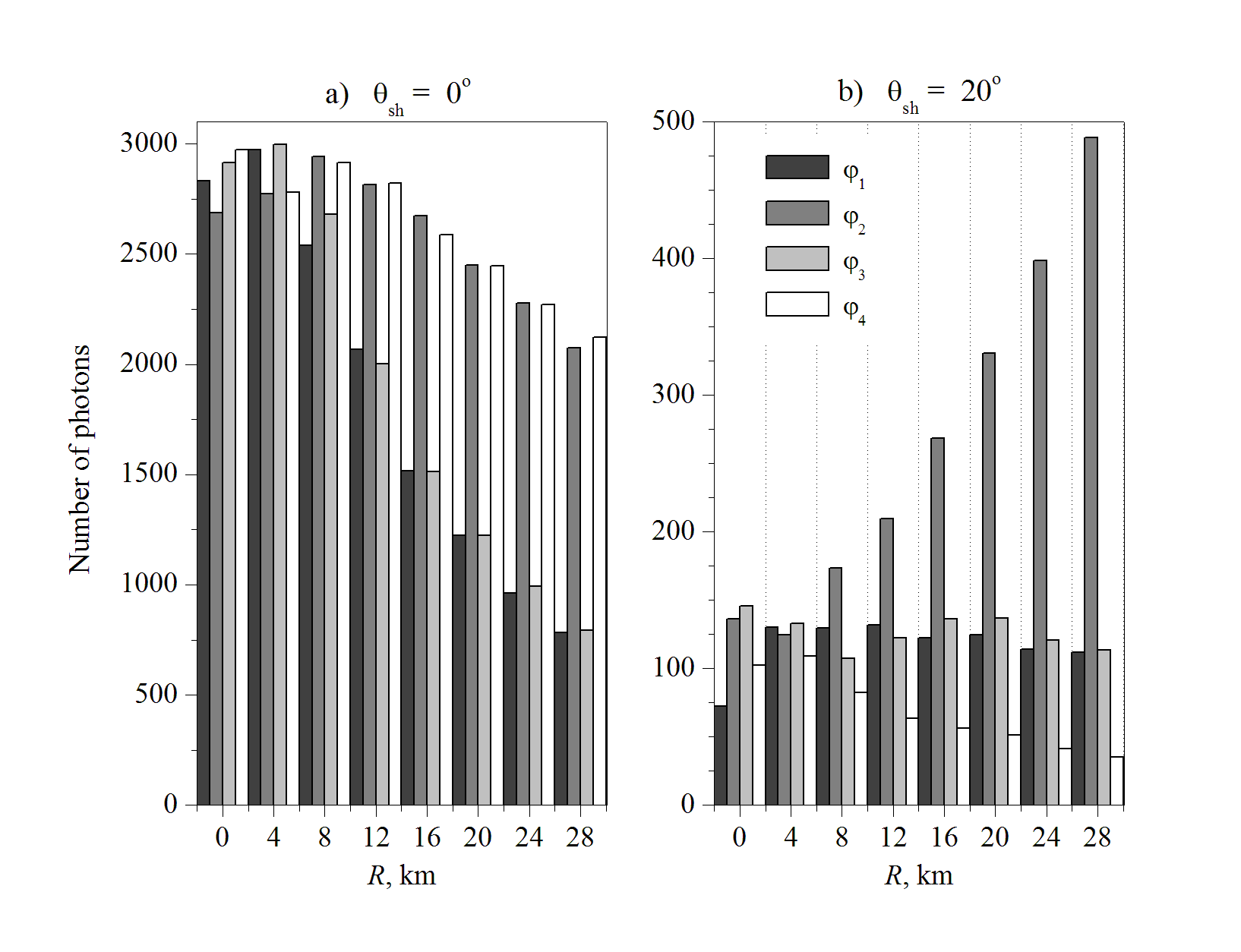}
\end{center}
\vspace{-1.5pc}
\caption{Number of ref{}lected Cherenkov photons that hit the detector mirror, depending on a value of $R$ for showers with zenith angles $\theta_\mathrm{sh}\!=\!0^\circ$\,(a) and $20^\circ$\,(b). The area with radius $R\!=\!30$\,km at the height of 400\,km is divided into four sectors on which the photons impinge with azimuth angles within the following ranges:  $\varphi_1\!=\!45^\circ-135^\circ$, $\varphi_2\!=\!135^\circ-225^\circ$, $\varphi_3\!=\!225^\circ-315^\circ$, $\varphi_4\!=\!315^\circ-45^\circ$.}\label{pic2}
\end{figure}

Let the detector to be located right above the origin of the laboratory system, at the point $R\!=\!0$. Here $R$ characterizes the location of the detector and is defined as $\sqrt{x^2+y^2}$ at a height of $z\!=\!400$\,km. Then the central PEM cell scans the area where $\approx\!96$\,\% of ref{}lected Cherenkov photons are concentrated. It means that practically all signal will be detected by a single cell of the PEM matrix. But only photons with zenith angles $\lesssim\!0.29^\circ$ will be focused on it. Our calculations show that at the height of 400\,km these photons almost uniformly cover a circle with radius $R\!\approx\!2$\,km. Therefore the signal can be estimated as product of the photon surface density in the circle and the mirror area of the detector, which yields $\approx\!3\!\cdot\!10^3$. The indicated number of ref{}lected photons will hit the mirror and will be focused on the central cell if coordinates of the detector satisfy the conditions:
\begin{equation*}
\sqrt{x^2_\mathrm{det}+y^2_\mathrm{det}}\leq2\;\text{km},\;\; z_\mathrm{det}=400\;\text{km}.
\end{equation*}

If the detector is located within the ring, $R\!=\!2\!-\!6$\,km, the area where almost all signal from the ref{}lected Cherenkov component is concentrated, is scanned by the cell bordering on the central one. Only the photons with zenith angles $\theta\!\approx\!(5\!-\!10)\!\cdot\!10^{-3}$\,rad will be focused on it. The number of the photons that will hit the mirror of the TUS detector, depending on its location, is presented in Fig.\,\ref{pic2}.

\begin{figure}[h!]
\begin{center}
\includegraphics[height=23pc]{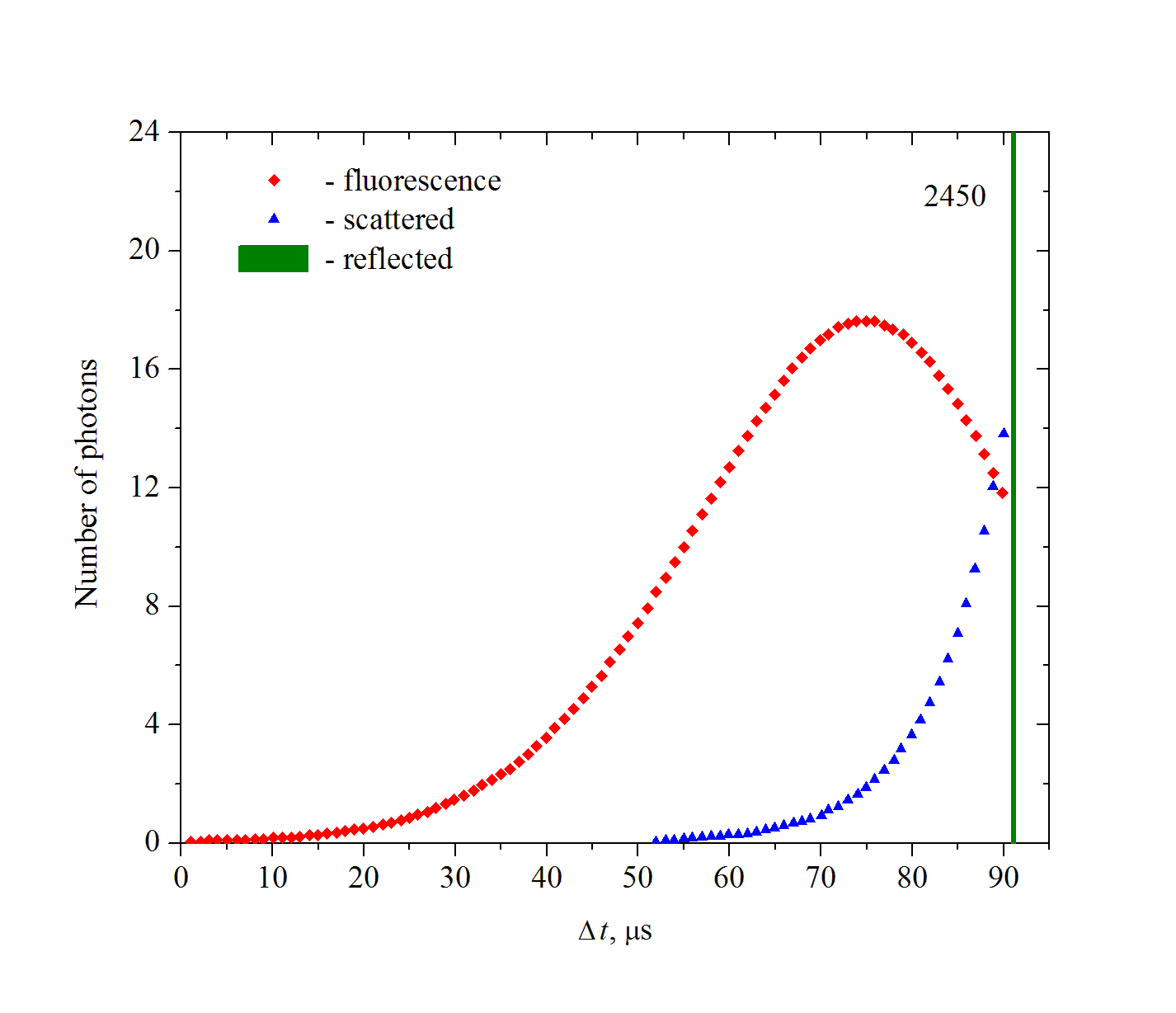}
\end{center}
\vspace{-1.5pc}
\caption{Time-base sweep of the signal recorded by the TUS detector at the distance $R\!=\!20$\,km for a vertical shower. The number of the ref{}lected photons in the f{}lash is also given. The time interval $\Delta t$ starting from the moment when the shower passes a depth of $200$\,g/cm$^2$, is plotted on the abscissa axis.}\label{pic3}
\end{figure}

Besides angular distribution of the signal, it is necessary to consider its duration as well. Calculations indicate that the duration of the ref{}lected signal is small as compared with the signal durations of the f{}luorescence and scattered Cherenkov components. Moreover, whereas the time resolution of the TUS detector is roughly equal to 1\,$\mu$s, we may consider all ref{}lected Cherenkov photons to be recorded simultaneously. Time-base sweep of signals from a vertical EAS, recorded by the TUS detector at the distance $R\!=\!20$\,km, is shown in Fig.\,\ref{pic3}. The f{}luorescence signal circumscribes the cascade curve of the shower and lasts $\approx\!100\,\mu$s. The signal from the scattered Cherenkov component is superimposed on the former one and blurs its maximum. This circumstance should be taken into account in processing experimental data. A short bright f{}lash of the ref{}lected Cherenkov photons is observed at the end of the f{}luorescence track, which is evidence of the shower entering into the ocean.

\vspace{-1em}

To clarify the question at what zenith angles $\theta_\mathrm{sh}$ of EASs the registration of the ref{}lected signal by the TUS detector is possible, let us look at Fig.\,\ref{pic1}(b) where the angular distribution of the ref{}lected Cherenkov photons from a shower with $\theta_\mathrm{sh}\!=\!20^\circ$ is presented. Obviously, the surface density of the photons with zenith angles up to $4.75^\circ$ ($r\!\lesssim\!0.083$) is substantially smaller than that from a vertical EAS. Additionaly, in our model, zero azimuth angle of the shower means that the shower axis lies in the plane perpendicular to the wave front. The situation will change if $\varphi_\mathrm{sh}$ dif{}fers from $0^\circ$ and $180^\circ$. According to our calculations, the TUS detector will not record with confidence the ref{}lected signal from a shower with angles $\theta_\mathrm{sh}\!=\!20^\circ$ and $\varphi_\mathrm{sh}\!=\!90^\circ$.
Therefore we restrict the range of zenith angles by a value of $20^\circ$.

\mysecv{Conclusions}

The calculations performed show that the main contribution to the Cherenkov radiation that occurs in developing of an EAS over the ocean is made by photons ref{}lected from the water surface and scattered in the atmosphere. Registration of the ref{}lected photons at the end of the f{}luorescence track may help us to determine the shower trajectory more precisely. However, the TUS detector, with the field of view $\approx\!9.5^\circ$, will be able to identify similar flashes only in showers with zenith angles up to $20^\circ$. The JEM-EUSO project is assumed to extend this range.

\mysecv{Acknowledgments}

This work was supported by Russian Foundation for Basic Research (grant 09-02-12162-ofi\_m).

\end{document}